# A Diffuse Light Field Imaging Model for Forward-Scattering Photon-Coded Signal Retrieval

Hongkun Cao, Xin Jin, *Senior Member, IEEE*, Junjie Wei, Yihui Fan and Dongyu Du

*Abstract*—Scattering imaging is often hindered by extremely low signal-to-noise ratios (SNRs) due to the prevalence of scattering noise. Light field imaging has been shown to be effective in suppressing noise and collect more ballistic photons as signals. However, to overcome the SNR limit in super-strong scattering environments, even with light field framework, only rare ballistic signals are insufficient. Inspired by radiative transfer theory, we propose a diffuse light field imaging model (DLIM) that leverages light field imaging to retrieve forward-scattered photons as signals to overcome the challenges of low-SNR imaging caused by super-strong scattering environments. This model aims to recover the ballistic photon signal as a source term from forward-scattered photons based on diffusion equations. The DLIM consists of two main processes: radiance modeling and diffusion light-field approximation. Radiate modeling analyzes the radiance distribution in scattering light field images using a proposed three-plane parameterization, which solves a 4-D radiate kernel describing the impulse function of scattering light field. Then, the scattering light field images synthesize a diffuse source satisfying the diffusion equation governing forward scattering photons, solved under Neumann boundary conditions in imaging space. This is the first physically-aware scattering light field imaging model, extending the conventional light field imaging framework from free space into diffuse space. The extensive experiments confirm that the DLIM can reconstruct the target objects even when scattering light field images are reduced as random noise at extremely low SNRs. Compared to state-of-the-art scattering light field imaging method like Peplography, the proposed method outperforms Peplography by 1.70dB/4.76dB Peak-Signal-Noise-Ratio (PSNR) and 0.167/0.172 Structural-Similarity-Index-Measure (SSIM) higher, on average, for passive-luminous/self-luminous targets, respectively. Facing experimental scattering images, most of dehaze and under water imaging methods fail to recover target object shape even after multi-view superimpose.

## I. Introduction

Scattering imaging is beneficial to many fields, such as auto-driving, underwater exploration, aviation safety, and so on. Unfortunately, imaging through highly scattering media is challenging because the propagation paths for most of the light are random walks, resulting in noise in the image, only the slight ballistic photons turn into signals. Scattering imaging has two primary subfields that are relevant to realistic applications in machine vision: dehazing and underwater imaging. Both are founded on identical fundamental principles of imaging formation, but with specific adjustments made to accommodate the distinct physical properties present under air and water in terms of light and medium. At a basic level, besides the ballistic photons, the scattered light consists of forward-scattered photons reflected by objects and backward-scattered photons that are solely reflected and refracted by scattering media. The McGlamery-Jaffe model defines these photon components as direct transmission photons, backward photons, and forward photons, which can be linearly combined together. By neglecting forward-scattering photons, the J-M model can be simplified to include only direct transmission light and atmospheric light [3-5]. Building upon this simplified J-M model, various dehazing and underwater imaging methods have been proposed [3-24]. Image restoration techniques employ prior knowledge to invert the degradation process and estimate parameters to obtain clear images [5]. A classic approach, the dark channel prior (DCP) [6-9], has been successfully applied to foggy image restoration. Furthermore, a range of other priors have been proposed, including the attenuation curve prior [10],

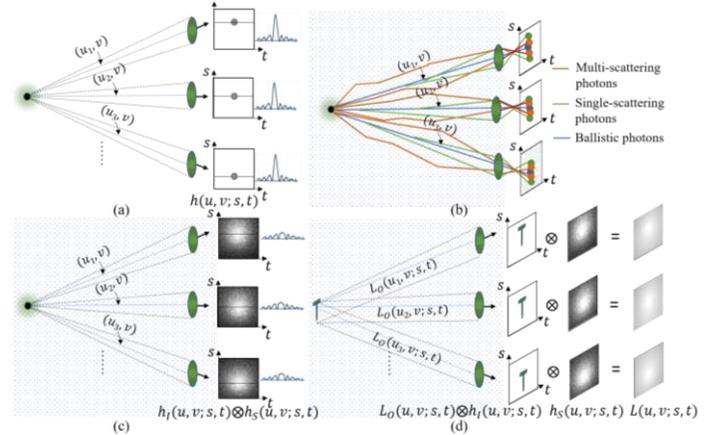

Fig. 1. The construction principle of light field imaging in scattering media. (a) The PSF of conventional light field imaging without considering scattering effect. (b) The scattering light field images consist of various scattering photons. (c) The PSF of scattering light field images. (d) The scattering light field imaging framework that is represented by the convolution between the object image and scattering PSF.

red color prior [11], statistical prior [12], haze line prior [13,14], illumination channel sparsity prior [15], Gamma correction prior [16], color attenuation prior [17] and extreme channel prior [18]. The image enhancement methods, such as Fusion

This work was supported in part by National Natural Science Foundation of China (NSFC) under Grants 62205178, 61991451, and in part by Shenzhen Science and Technology Project under Grant (JSGG20210802154807022), and The Major Key Project of PCL (PCL2023A010). (*Corresponding author: Xin Jin*).

Hongkun Cao is with in Peng Cheng Laboratory, Shenzhen 518055, China (e-mail: caohk@pcl.ac.cn ), Xin Jin, Junjie Wei , Yihui Fan and Dongyu Du are with the Shenzhen Key Laboratory of Broadband Network and Multimedia, Shenzhen International Graduate School, Tsinghua University, Shenzhen 518055, China (jin.xin@sz.tsinghua.edu.cn; weijj21@mails.tsinghua.edu.cn; e-mail; fyh20@mails.tsinghua.edu.cn; dudy19@mails.tsinghua.edu.cn ).

[19],[20], Retinex [21],[22] addressing the signal only in image space, which significantly improves the image contrast. Recently, the spectral-based underwater image process tackle high- and low-frequency components of scattering images separately and fuse them using specific transforms [23]. While data-driven methods have garnered immense attention [24-27], obtaining high-quality scattering data for foggy and underwater environments proves to be a challenging task due to their inherent complexity.

Despite the significant success achieved by these methods in dehazing and underwater imaging, their capabilities are limited at dealing with thin scattering areas. This limitation arises from the fact that the effective signal provided solely by ballistic photons decreases as scattering intensifies, even after accounting for compression of scattering noise. As a result, the signal-to-noise ratio (SNR) of the reconstructed image remains low, and it is effective that using designed capturing system with larger aperture, dynamic range, etc. to achieve SNR gains further and then develop corresponding algorithm to process the synthesized image, which counterbalances the lack of effectiveness of single image underwater or dehazing processing algorithms.

One promising method is using multiple views to capture lights with more signal collection and noise suppression, increasing the SNR of final synthesized images [28-35]. For instance, light field imaging methods aim to distinguish ballistic photons from scattering photons by using the intensity consistency of direct light in all perspective images [28, 29] and extract signal by assuming the Gaussian distribution of scattering lights [30], as well as the depth extraction method [31]. Peplography models scattering lights with statistical estimation and extract the direct light based on the photon counting model under a light field imaging framework [32].

As we know, the forward scattering photon utilization has not been mentioned except for the work by Pan. et [11], as the small ratio compared to other photon components and the difficulty to restore them into signal. In the light field imaging model, the function of forward scattering photons can be amplified, since total amount are increased after accumulating all views and the implicit object information can be meaningful if we can effectively use them. Here, a physically-aware imaging model is derived leveraging the light field framework based on radiate transfer theory, which is called diffuse light field imaging model (DLIM). Beyond ballistic photon signal, it is capable of converting the forward scattering photons into signal by solving diffusion equation (DE). DLIM consists of two main processes: radiate modeling and diffusion approximation. Radiate modeling analyzes the radiance distribution in scattering light field images with a proposed three-plane parameterization, which solves a 4-D radiate kernel as reconstruction filter for scattering light field. Then the scattering light field images synthesize a diffuse source satisfying the DE ruling the forward scattering photons with diffusion approximation, which is solved under robin boundary condition in imaging space. Finally, the object image can be reconstructed from the refocused scattering light field image using the diffuse kernel with computational complexity $O(N^2 log N)$ for $N \times N$ image resolution. In addition, the backscattering photons (atmosphere light) still can be removed based on J-M model. Compared with state-of-the-art scattering light field imaging method Peplography [32], the proposed SLIM outperforms Peplography by 1.70dB/4.76dB Peak-Signal-Noise-Ratio (PSNR) and 0.167/0.172 Structural-Similarity-Index-Measure (SSIM) higher, on average, for passive-luminous/self-luminous targets, respectively. Due to lack of scattering light field imaging method, all kinds of single image dehazing methods are also investigated as comparison, in which all view images are superimposed for denoising and then processed by those methods. The results confirm that the DLIM still gain highest PSNR values of reconstructed images.

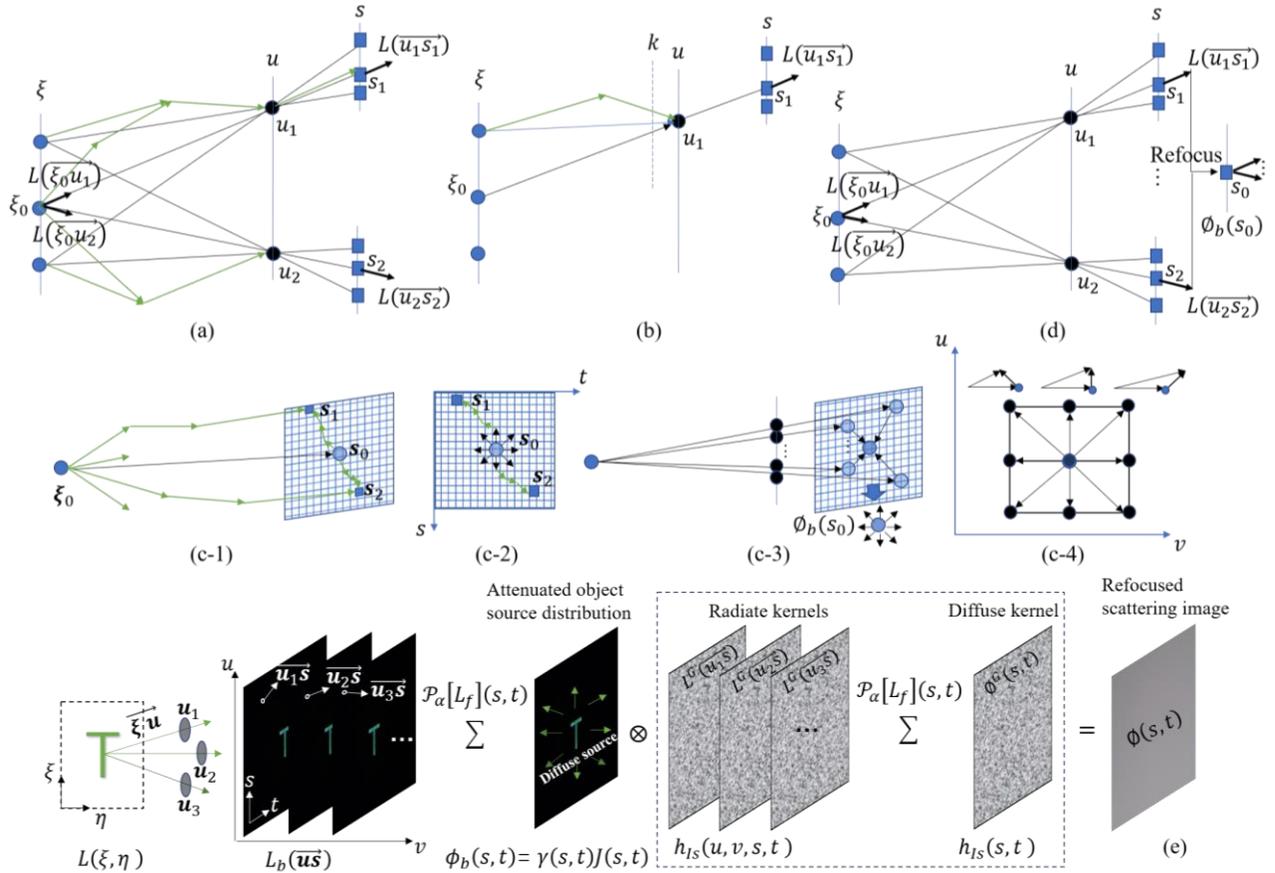

Fig.2. The schematic illustration of physical mechanism for diffuse light field imaging. (a) Analysis of radiance distribution in light field images considering both scattering and ballistic photons. (b) Example of modeling scattering photons contributing to $L(\overrightarrow{u_1 s_1})$. (c) The diffusion approximation derivation. (c-1) Approximation to photon diffuse route in three-dimensional space by two-dimensional diffuse in imaging plane. (c-2) Requirement of diffuse source to form diffuse process in imaging plane. (c-3) Construction of diffuse source in imaging plane taking advantage of light field containing ballistic photons from various directions. (c-4) The projections of transport line of ballistic photon of light field contribute concentric radiances as diffuse source in refocused image. (d) Example for refocus explanation and implementation. (e) Forward framework of diffuse light field imaging modeling (DLIM).

The remainder of this paper is organized as follows. Section II demonstrates the mechanism and framework of proposed diffuse light field imaging model. Section III conducts the experiment implementation and analyzes the experimental results. The discussions and conclusions are provided in Section IV.

## II. PROPOSED DIFFUSE LIGHT FILED IMAGING MODEL

As shown in Fig. 1(a), in a linear light field imaging system, a quasi-point of impulse response can be observed in the imaging plane for a conventional imaging system, and the images can be represented by the convolution between light source distribution and point spread function (PSF) as follows:

$$L(u,v,s,t) = L_O(u,v,\xi,\eta) \otimes h_I(u,v,s,t), \quad (1)$$

where $L(u,v,s,t)$ is the light field images, $L_O(u,v,\xi,\eta)$ and $h_I(u,v;s,t)$ represent the distribution of the light source and the point spread function (PSF) of imaging system, respectively. In a scattering light field imaging, the ballistic photons can penetrate scattering media as depicted by blue points, while the other two types of scattering photons are randomly distributed in the imaging plane as depicted by red and green points as shown in Fig. 1(b). As a result, the impulse response, in other words, the PSF of scattering light field becomes speckle patterns, in which tiny highlights of intensities in images near ballistic-photon accumulated positions can be observed as seen in Fig. 1(c). Most highly scattering medias, such as dense fog, turbid water, etc., are uniform and homogeneous, which still could be regarded as a linear imaging system if we scarify some imaging resolution. With this approximation, another convolution term as scattering impulse denoted by $h_S(u,v,s,t)$ is introduced into the imaging process as defined in Eq. (2), as depicted by Fig. 1(d).

$$L(u,v,s,t) = L_O(u,v,\xi,\eta) \otimes h_I(u,v,s,t) \otimes h_S(u,v,s,t) \quad (2)$$

### A. Three-plane parameterization

To derive the scattering light field imaging model conveniently, a three-plane coordinate representation is designed. The conventional two-plane parameterization only add another angular dimension to spatial dimension in imaging space to record the emitting direction of light from object. This is under the condition that the direction of $\mathbf{u}(u,v) \rightarrow \mathbf{s}(s,t)$ is equal to the emitting direction of light from object to camera, since only line transport of ballistic photons is involved. However, in the diffuse light-field framework, the direction of $\mathbf{u}(u,v) \rightarrow \mathbf{s}(s,t)$ is not equal to the emitting direction of light from object to camera, since not only ballistic but also scattering photons came into $\mathbf{u}$-plane and received by $\mathbf{s}$-plane. Yet it is critical to derive the correlation between ballistic and scattering photon in our modeling process. Thus, how to scientifically describe the radiances in spatial and angular dimensions of both ballistic and scattering photons under light field framework is important. In the three-plane parameterization, we add another plane $\boldsymbol{\xi} = (\xi, \eta)$ in object space and using the two-plane of ($\boldsymbol{\xi}, \mathbf{u}$) specifically describe the radiances of ballistic photons with $L(\boldsymbol{\xi}, \mathbf{u})$, since those photons from plane $\boldsymbol{\xi}$ to plane $\mathbf{u}$ through line transport must be ballistic photons. The $L(\mathbf{u}, \mathbf{s})$ still represents scattering photon radiances. As seen in Fig. 2, for simplicity, only one dimension for each plane is considered for each plane and all derivations are appropriate for the two-dimensional plane. For example, the radiance of the object point $\xi_0$ towards $u_1$ is denoted as $L(\overrightarrow{\xi_0 u_1})$, and the radiance received in pixel $s_1$ coming from $u_1$ is denoted as $L(\overrightarrow{u_1 s_1})$, as seen in Fig. 2 (a). Noting that the $L(\overrightarrow{\xi_0 u_1})$ is only ballistic photon, whereas the $L(\overrightarrow{u_1 s_1})$ could be ballistic, scattering photons and both of them. In particular, we add another auxiliary plane between object plane and camera plane to describe the radiances of incident scattering photons came to camera plane for constructing radiate transfer function. Then the key component is how to construct the relation of radiances between those ballistic photons, scattering photons under above three-plane parameterization, which is called radiate modeling.

### B. Radiate modeling

Radiate modeling refers to analyzing the radiance distribution in scattering light field images with three-plane parameterization. Here we take only one perspective $u_1$ of the light field as an example. The received radiance is contributed by two terms composed of ballistic photons and scattering photons as defined by

$$L(\overrightarrow{u_1 s}) = L_b(\overrightarrow{u_1 s}) + L_s(\overrightarrow{u_1 s}), \quad (3)$$

where $L_b(\overrightarrow{u_1 s})$ indicates the ballistic term and $L_s(\overrightarrow{u_1 s})$ means the scattering term as depicted by black and green line in Fig. 2 (a)[35].

The most inspiring idea in the whole modeling is to establish the correlation between scattering and ballistic photons, so that the ballistic photons carrying object source information can be recovered from scattering photons. Here the ballistic photons are attenuated linearly as the distance between camera and object increases, as defined by

$$L_b(\overrightarrow{u_1 s}) = L(\overrightarrow{\xi u_1}) \exp(-\mu_s |\overrightarrow{\xi u_1}|), \quad (4)$$

where $\mu_s$ is the scattering coefficient and the absorption coefficient has been neglected here considering weak absorption of scattering media. The scattering photons $L_s(\overrightarrow{u_1 s})$ consists of single and multiple scattering photons as depicted by blue and green lines between $\xi$ and $\mathbf{u}$ planes in Fig. 2(b), can be mathematically described as

$$L_s(\overrightarrow{u_1 s}) = \mu_s \int L(\overrightarrow{ku_1}) p(\overrightarrow{ku_1}, \overrightarrow{u_1 s}) dk, \quad (5)$$

where $k$ means the coordinate in the auxiliary plane before camera plane as seen in Fig. 2(b) and $p(\overrightarrow{ku_1}, \overrightarrow{u_1 s})$ is the phase function. Thus, based on energy conservation, a partial differential equation can be constructed as

$$\overrightarrow{u_1 s} \cdot \nabla L(\overrightarrow{u_1 s}) + \mu_{eff} L(\overrightarrow{u_1 s}) = L_b(\overrightarrow{u_1 s}) + L_s(\overrightarrow{u_1 s}). \quad (6)$$

The $\mu_{eff} = \mu_a + \mu_s'$ is the effective attenuation coefficient, where $\mu_a$ and $\mu_s'$ are the absorption and reduced scattering coefficients [35].

It is revealed that the received radiances are correlated to the object radiance through source term $L_b(\overrightarrow{u_1 s})$ of above radiate transfer equation (RTE), since $L_b(\overrightarrow{u_1 s})$ is linearly attenuation version of the object radiance $L(\overrightarrow{\xi u_1})$ according to Eq. (4). Thus, if the source term $L_b(\overrightarrow{u_1 s})$ can be reconstructed, the object radiance will be obtained as a result. To solve above equation, the Green function can be introduced. Let's assuming an impulse object source, such as $\delta(|\overrightarrow{\xi u_1}| - |\overrightarrow{\xi_0 u_1}|)$, then $L_b^\delta(\overrightarrow{u_1 s}) = L\left(\delta(|\overrightarrow{\xi u_1}| - |\overrightarrow{\xi_0 u_1}|)\right) \exp(-\mu_s |\overrightarrow{\xi_0 u_1}|)$. Let $L_b(\overrightarrow{u_1 s}) = L_b^\delta(\overrightarrow{u_1 s})$ in Eq. (6), the solution is an impulse response of this system, which is called Green function and denoted as $L^G(\overrightarrow{u_1 s})$. Then the radiance distribution in imaging plane of perspective $u_1$ can be modeled as convolution between ballistic radiance distribution $L_b(\overrightarrow{u_1 s})$ and Green function $L^G(\overrightarrow{u_1 s})$ as

$$L(\overrightarrow{u_1 s}) = L_b(\overrightarrow{u_1 s}) \otimes L^G(\overrightarrow{u_1 s}). \quad (7)$$

The Green function can be solved by expanding radiance with spherical harmonics, an infinite approximation to RTE is obtained. The $P_N$ approximation means taking the first $N$ spherical harmonics, which gives $(N+1)^2$ coupled partial differential equations. Diffusion approximation as the $P^1$ approximation to RTE has wide usage in biological imaging [36-37]. When imaging distance larger than $\ell_t$ that is transport mean free path and the media is isotropic, the radiate transfer is well described by diffusion approximation [37-38].

### C. Diffusion light-field approximation

According to diffusion approximation [35], we take the integral along angular dimension of scattering light field, e.g. the $\mathbf{u}$ dimension under Eq. (6), then a diffusion equation is constructed as

$$D\nabla^2 \emptyset(s) - \mu_a \emptyset(s) = \emptyset_b(s), \quad (8)$$

where $D$ is the diffuse coefficient $1/(3*(\mu_a + \mu_s(1-g)))$, $g$ is the isotropic coefficient [35]. Then the radiances are integrated as fluence rate noted by $\emptyset(s) = \int L(\overrightarrow{us}) du$ [35]. Noting that a diffuse sourced term is formed by integral as $\emptyset_b(s) = \int L_b(\overrightarrow{us}) du = \gamma \int L(\overrightarrow{\xi u}) du$, where $\gamma = \exp(-\mu_s |\overrightarrow{\xi u}|) \approx \exp(-\mu_s z)$ is the Lambert-Beer attenuation ratio. If we consider the depth of object sources, the diffuse source can be defined as $\emptyset_b(s) = \emptyset(\xi)\gamma(s)$, where coordinates $\xi$ and $s$ have been matched by imaging system and $\gamma(s) = \exp(-\mu_s z(s))$.

Here we use the refocus operation of light field imaging to approximate the above diffuse approximation, i.e. the angular integral. The physical mechanism is demonstrated as seen in Fig. 2 (c). Assuming a point light source at $\xi_0$, it will radiate photons to various directions and the scattered photons reach pixels of imaging plane randomly, such as $\mathbf{s_1} = [s_1, t_1; z+f]$, $\mathbf{s_2} = [s_2, t_2; z+f]$, where $f$ is focusing length of imaging lens. As seen in Fig. 2(c-1), the photon diffuse route from source to sensor pixel ($\xi_0 \to s_1$), $\xi_0 = [\xi_0, \eta_0; 0]$, could be approximated by two parts containing line transport of ballistic photon ($\xi_0 \to s_0$), $s_0 = [s_0, t_0; z]$, and the diffusion in imaging plane ($s_0 \to s_1$), same as other pixels, such as $s_2$. This is enabled under a precondition that a diffuse source exists in imaging plane, which allows all pixels in imaging plane can be diffused by the source as seen in Fig. 2(c-2). In the light-field approximation model, the diffuse source is formed by accumulating radiances coming from various directional ballistic photons of object source leveraging multi-view capturing. That is, taking advantage of light field imaging framework, the ballistic photons of object source can be captured by various perspectives from different directions and all directional object radiances will be integrated by refocus and then a diffuse source $\emptyset_b(s_0)$ is constructed in refocused image of scattering light field as seen in Fig. 2(c-3), as a result the whole refocused image can be the diffuse result of this source term. More specifically, assuming a grid of capture spots as shown in Fig. 2(c-4), each projection of the lines of ballistic photon transport from the object source to each spot will contribute one concentric radiance to the diffuse source.

For instance, an object point source $\xi_0$ omits various directional radiances, such as $\overrightarrow{\xi_0 u_1}$ and $\overrightarrow{\xi_0 u_2}$, will be integrated at pixel $s_0$ of refocused image as seen in Fig. 2(d). Then the attenuated version of $\phi(\xi_0)$ in imaging plane is denoted as $\phi_b(s_0)$ that is defined as

$$\emptyset_b(s_0) = \sum_{i=1}^{M} L_b\left(\overrightarrow{u_i(s_i - \iota\Delta s)}\right), \quad (9)$$

where $i$ denotes the perspective index and $M$ is the number of perspectives, $\Delta s$ is one-time refocus shift distance in imaging plane corresponding to the baseline of camera array. According to Eq. (4), $L_b\left(\overrightarrow{u_i(s_i - \iota\Delta s)}\right) = \gamma L(\overrightarrow{\xi_0 u_i})$, which corresponding to the part of line transport of ballistic photons.

The solution of the Eq.(8) assuming impulse object source, such as $\delta(\xi - \xi_0)$, is the Green function $\emptyset^G(s)$, which is determined by setting of $\emptyset_b(s) = \emptyset_b^\delta(s)$, where $\emptyset_b^\delta(s) = \emptyset(\delta(\xi - \xi_0))\gamma$. In this way, the diffuse process in the angular-integrated light-field imaging space can be represented as convolution between the diffuse source distribution and the Green function as defined in Eq. (10), where the former is the attenuated version of the fluence-rate distribution of the object source.

$$\emptyset(s) = \emptyset_b(s) \otimes \emptyset^G(s) \quad (10)$$

In order to solve the Green function of DE analytically, the boundary conditions of the media involved must be determined. If we consider the media as dense fog mainly composed of water and the camera lens as a glassy material, the refractive gradient between the fog and the lens is neglected. We can use the Neumann boundary ($\frac{\partial \emptyset}{\partial \vec{r}} = 0$) to solve Eq. (8) [38, 39]. The mathematical procedure for solving DE can be found in any literatures refer radiate transfer. Here we give the result directly, which is defined as

$$\emptyset^G(s) = \frac{1}{2\sqrt{\mu_a D}} \left( e^{-\sqrt{\frac{\mu_a}{D}}|s - s_0|} + e^{-\sqrt{\frac{\mu_a}{D}}|s + s_0|} \right) \quad (11).$$

Let's get back to the original light field imaging equation of Eq. (2), $h_I(u,v,s,t) \otimes h_S(u,v,s,t)$ is combined as radiate kernel $h_{IS}(u,v,s,t)$ according to associate law of convolution, which can be obtained by solving the radiate Green function $L^G(\overrightarrow{us})$, $\boldsymbol{u} = [u,v;z]$, $\boldsymbol{s} = [s,t;z+f]$. According to the theorem of filtered light field photography, the 4-D convolution of light field can be simplified as 2-D convolution, as defined

$$\mathcal{P}_\alpha \circ \mathcal{C}^4_\mathcal{K} \equiv \mathcal{C}^2_{\mathcal{P}_\alpha[\mathcal{K}]} \circ \mathcal{P}_\alpha \quad (12)$$

where $\mathcal{P}_\alpha$ is the refocus operator defined as

$$\mathcal{P}_\alpha[L_f](s,t) = \frac{1}{\alpha^2 f^2} \iint L_f\left(u\left(1-\frac{1}{\alpha}\right) + \frac{s}{\alpha}, v\left(1-\frac{1}{\alpha}\right) + \frac{t}{\alpha}, u, v\right) du dv \quad (13),$$

where $\alpha = \frac{f}{f+z}$. The $\mathcal{C}^N_\mathcal{K}$ is an N-dimensional convolution operator with filter kernel $\mathcal{K}$[40]. With the derived diffusion light field imaging model, this theorem is also applicable by setting the integral of 4D filter kernel as a diffuse kernel, i.e. $\mathcal{P}_\alpha[\mathcal{K}] = \emptyset^G(s)$, when imaging through scattering media and the diffuse kernel $\emptyset^G(s)$ is intrinsically equal to $\mathcal{P}_\alpha[L^G(\overrightarrow{us})]$.

The above forward process is illustrated heuristically by the diagram in Fig. 2(e). The radiances of an object with the source distribution $L(\xi,\eta)$ are received by the sensors through different lenses at ($\boldsymbol{u}_1$, $\boldsymbol{u}_2$, $\boldsymbol{u}_3$, ...). The received radiances are represented as radiate source convolving radiate kernel ($L_b(\overrightarrow{us}) \otimes L^G(\overrightarrow{us})$). Leveraging theorem of filtered light field photography, we can refocus $L_b(\overrightarrow{us})$ and $L^G(\overrightarrow{us})$ separately, so that a diffuse source $\phi_b(s,t) = \gamma(s,t)J(s,t)$ and diffuse kernel $\emptyset^G(s,t)$ can be obtained. By the way, $J(s,t)$ can be clear image of the object source without scattering. Most importantly, with the diffusion approximation, the diffuse kernel is directly solved based on the diffusion equation and the complicated calculation to radiate kernel is avoided. Finally, the refocused scattering light field image $\phi(s,t)$ can be represented as an implicit object source distribution $\phi_b(s)$ convolving the diffuse kernel $\emptyset^G(s)$.

Backward-scattering modeling is required for imaging in dealing with nature scattering environment, in which the light source is separated from object, and the back scattering photons without reaching object are involved as strong noise. In fact, the noise caused by backward scattering photons can be random noise, taking advantage of light field imaging framework, the random noises can be suppressed by superimposing each other. Please see the detailed derivation in Supplementary.

Here, we combine the proposed DLIM and simplified J-M model, noted as DLIM-J to account for forward scattering lights, backscattering lights, ballistic lights, and light source distribution under light field imaging framework as defined by

$$J^*(s,t) = \phi_b(s,t) \otimes (\emptyset^G(s,t) + 1) + B_\infty(1 - \gamma(s,t)), \quad (14)$$

where $J^*(s,t)$ is the scattering image, $\phi_b(s,t) = \gamma(s,t)J(s,t)$ is the distribution of attenuated object source, $J(s,t)$ is the clear image, the ballistic attenuation ratio $\gamma(s,t)$ is the same as the medium transmission $t(s,t)$ in J-M model and $B_\infty$ indicates the atmosphere lights, respectively. Here the wavelength is not discussed in whole model derivation, although it is also correlated with the scattering effect, but is out of the scope of this research.

### D. Inversion procedure

The recovery from the scattering image $J^*(s,t)$ to original image $J(s,t)$ is an inversion procedure of DLIM, which is accomplished by deconvolution using diffuse kernel $\emptyset^G(s,t)$. A closed-form solution exists using the Wiener deconvolution filter:

$$\hat{J} = \mathbf{F}^{-1} \circ \left[\frac{\Phi^G}{|\Phi^G|^2 + \frac{1}{\zeta}}\right] \circ \mathbf{F} \circ J^*. \quad (15)$$

$\mathbf{F}$ denotes the discrete Fourier transform matrix, $\Phi^G$ is the diagonal matrix whose elements correspond to the Fourier coefficients of the diffuse kernel $\emptyset^G$, $\zeta$ is a parameter that varies depending on the signal-to-noise ratio at each frequency, and $\hat{J}$ is estimated original image. The computational complexity of the above equation is only $O(N^2 \log N)$ for $N \times N$ image resolution. Noting that the self-luminous object will not require modeling backscattering photons. For example, various LED panels for advertisement at night, vehicle lights, diffuse optical tomography (DOT) in biological imaging, optical fiber imaging, e.g., whose scattering images are contributed only by forward scattering and ballistic photons. Notably, the optical thickness can be estimated using visibility multiplying $|log0.05|$, which can be measured with a Nephelometer for fog environments [41]. Then the scattering coefficient can be evaluated from optical thickness and depth that can be obtained with laser distance meter for outdoor environments. The exponential of optical thickness can be regarded as attenuation ratio. Here, to verify the convolution framework enabled by diffuse light field imaging framework, a PSF study of scattering light field imaging system is conducted, which proved that the PSF of diffuse light field image has consistent curve with the calculated diffuse kernel $\emptyset^G$ according to Eq.(11). Please found the detail in Supplementary about PSF retrieve and diffuse kernel calculation.

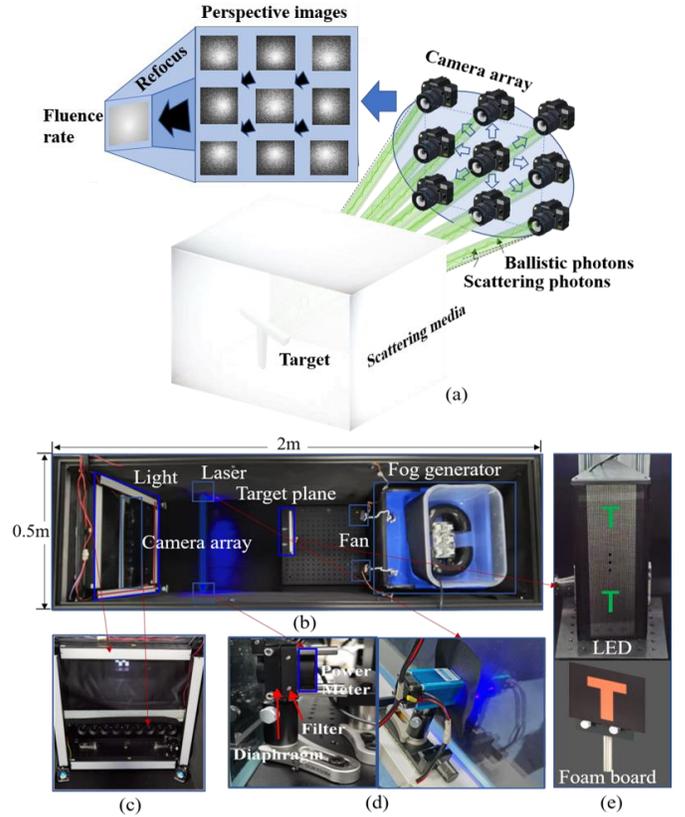

Fig. 3. The diagram of experimental system. (a) Experimental schematic of scattering light field imaging. (b) The experimental set-up. (c) Light and camera system. (d)The measurement system of optical thickness of scattering media. (e) The targets contain self-luminous LED light panel and passive-luminous target of printed foam board.

In the case of inversion procedure of backward scattering modeling, the Dark Channel Prior (DCP) as one of well-known J-M based algorithm is used to estimate the transmittance considering atmosphere light [5]. Then the recovered image by DCP will only contain forward scattering photons, noted as $J^*(s, t)$. More detail can be found in Supplementary.

scattering media that is built by a fog chamber, and the cameras are placed behind the scattering media in discrete capture positions. When calibrating the light field imaging system, the homography for each perspective is estimated by capturing checkboard images without filling the fog according to Zhang's method [42].

As shown in Fig. 3(b), the fog chamber mainly consists of an ultrasonic fog generator, a light and camera system, a measurement system of optical thickness, and the targets. The camera system is mounted outside the fog chamber. It consists of 9 cameras in a row and all cameras are connected to a synchronous signal generator to ensure the synchronous

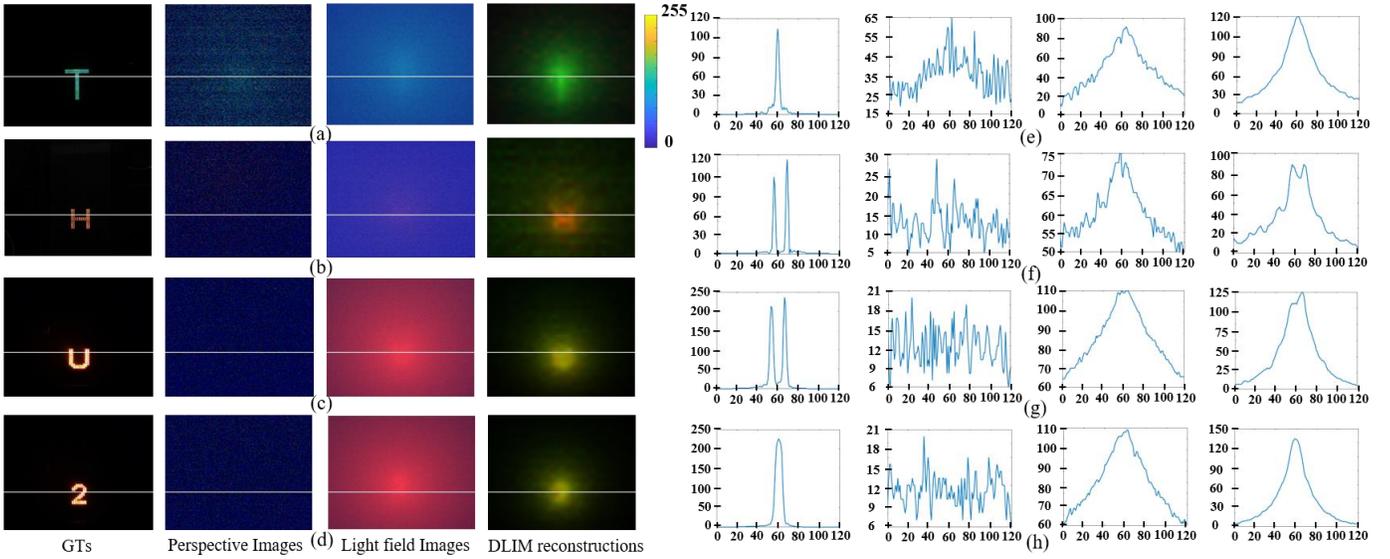

Fig. 4. Experimental results of verification for self-luminous targets by using DLIM. The 1st column refers to self-luminous target images. The scattering images for each target are shown in 2nd column. The light field images are shown in 3rd column. The DLIM-reconstructed results of scattering light field images are shown in 4th column. Images for self-luminous targets of 'T', 'H', 'U', '2' are shown in (a)-(d), and the curves corresponding to one intensity line of left images are shown in (e)-(h), respectively.

### III. EXPERIMENTS AND RESULTS

To verify the effectiveness and demonstrate the imaging capability of the proposed DLIM, the imaging experiments for both self-luminous and passive-luminous objects through dense fog under light field imaging framework have been conducted. The self-luminous experiments aim to confirm the reconstruction capability of DLIM by utilizing forward scattering photons, whereas the passive-luminous experiments aim to confirm its effectiveness involving strong noise caused by back scattered photons. The experimental scheme is shown in Fig. 3(a), in which different angles of light emitted from the target object are captured by cameras from different views, which are synthesized into the final fluence rate image with the refocus operation. Moreover, in terms of several critical factors effecting diffuse light field modeling, the simulation experiments using physical-based rendering platform have also been carried out to accurately control the relative coefficient values and the experimental results are carefully analyzed.

#### A. *Execution of a comprehensive experimental setup*

In practical experiments, the targets emerge into volumetric triggering (see Fig. 3(c)). Each camera consists of a 4mm lens and a CCD (Flir BFS-PGE-16S2; pixel size, 3.45 µm × 3.45 µm; resolution, 1440 × 1080). The illumination consists of four independent strip light sources (BRD36030), which are set for experiments of passive-luminous light field imaging, and switched off for experiments of self-luminous light field imaging. The system for measuring the optical thickness of scattering media consists of a laser with a wavelength of 480nm, a power meter, a filter and apertures, as shown in Fig. 3(d). Among them, the filter and diaphragms are to filter out scattering photons and only ballistic photons can be collected by the power meter to make the measurement more accurate. The optical thickness can be defined as $T = log\left(\frac{P_o}{P_a}\right)$, where $P_o$ and $P_a$ are the original laser power without scattering and the attenuated laser power by the scattering effect of fog, respectively [41]. To test the imaging capacity of the proposed methods, four group of self-luminous and three passive-luminous light field images were acquired through dense fog. The targets for self- and passive- luminous images are shown in Fig. 3(e). The self-luminous targets are the capital letters 'T', 'H', 'U' and a number '2' displayed on the LED panel with an upward motion, and the acquisition frequency is set to same as the motion frequency of letters. The self-luminous light field images contain a total of 8 horizontal and 16 vertical perspectives. The passive-luminous light field images consist of only 9 horizontal perspectives, which allows the image to be captured in a single step. The optical thicknesses for self- and passive- luminous experiments are measured as approximately OT =10 and 9 respectively. Explicitly, the thickness of fog is 0.7 m which is measured from the target to the front of the camera lens. For the optical thickness measurement system, the distance from the laser source to the power meter is 0.5 m. The original value of the power meter is 0.339W, and the corresponding value for capturing after filling with fog is 5.63e-4W, so T = 6.4 correspondingly. Considering that the distance ratio between the target object and the camera compared with the one between the laser and the power meter is about 1.4, the final optical thickness is about 9. For self-luminous objects, the distance between the target object and the camera lens is about 0.8 m, which means that the distance ratio becomes about 1.6, so the OT equals 6.4×1.6 ≈ 10. Even in artificial experimental environments such as a fog chamber, the optical thickness measurement is not accurate due to the fluctuations of the fog density during image acquisition. The exposure time is set as around 33ms.

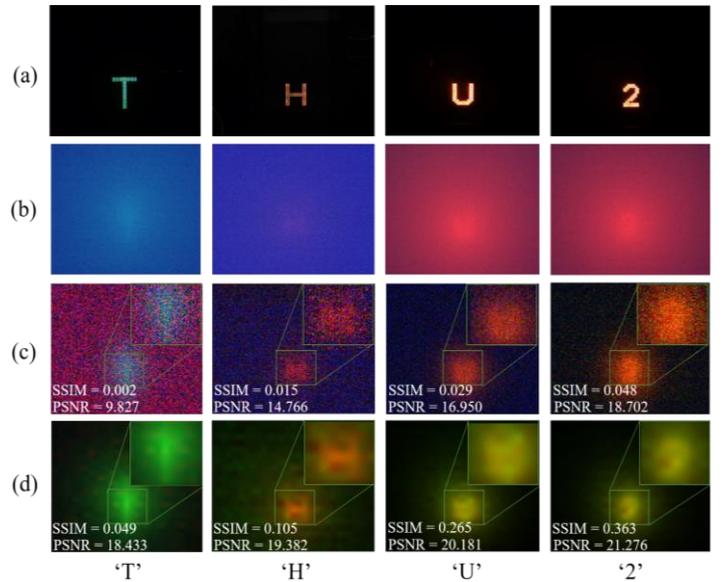

Fig. 5. Experimental comparison between Peplography and DLIM regarding self-luminous scattering light field images. (a) Ground truth. (b) Refocused scattering light field images. (c) Reconstructed images with Peplography. (d) Reconstructed images with proposed method.

*B. Verification of DLIM for self-luminous objects through super dense fog*

To verify the scattering imaging capacity of the proposed DLIM, 4 group experiments were performed for different self-luminous targets 'T', 'H', 'U' and '2' with different colors and intensities (see (a)-(d) in Fig (4)). As seen in Fig.4, the first column refers to self-luminous target images for the capital letters 'T', 'H', 'U' and the digit '2' as ground truth (GT). The second column shows one of the perspective images of scattering light field images that were normalized, and the third column shows the scattering light field images synthesized from $8_{horizontal} \times 16_{vertical}$ perspective images. And the fourth column shows the reconstructed results of scattering light field images for each target by using the proposed DLIM method. Note that the density of fog is hard to be controlled to be stable and consistent during capturing for each target, which will cause fluctuation of real optical thickness for different targets. From the captured perspective images, the differences of real scattering strength for different targets can be recognized, even if all scattering images for the targets are captured aiming at identical scattering conditions with the same optical thickness. The curves of intensity in one row of image are shown in (e)-(h) of Fig. 4. It is obvious that the single perspective image for all targets has a very low SNR, from which any target information is impossibly recovered. The curves of refocused light field images are all Gaussian converged, which is also consistent with our derived diffuse approximation. This shows that the diffuse source is indeed constructed in the refocused image and the whole image could be the result of source diffusion. The targets 'U' and '2' have higher intensity compared with 'T' and 'H', which aims to verify that if the improved intensity of the light source can enhance reconstruction quality. As seen in (c), (d) and (g), (h) of Fig. 4, the curves of intensity-improved perspective images are further shrunk, and a smoother curve is observed for the light field images, but the reconstruction quality of DLIM is not improved. This reveals that the slightly higher intensity of the light sources cannot improve the reconstruction quality of DLIM. This differs from ballistic-photon-based methods, whose imaging quality is directly correlated with the intensity of light source. All targets are successfully reconstructed from the scattering light field images using the proposed DLIM method. In contrast, the targets are impossibly resolved from refocused light field images directly, even though the speckle noise suppressed and ballistic photons have been accumulated.

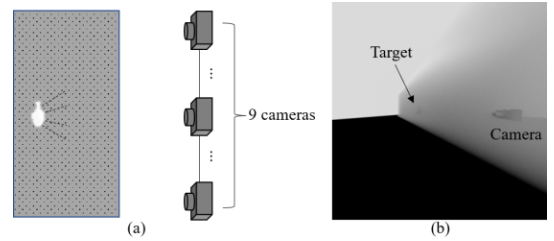

Fig. 6. Simulation diagram. (a) Scheme for simulation experiments. (b)The rendering simulation environment.

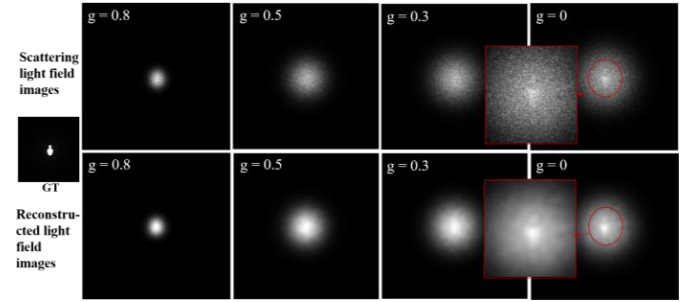

Fig. 7. Experimental results for verification of the effect of anisotropic factor of media to diffuse light field approximation.

Furthermore, the state-of-the-art scattering light-field imaging method, Peplography is implemented as a comparison [30]. In Peplography, the scattering photons are reduced using statistical modelling, and then the ballistic photons are accumulated using the photon counting [28]. The input images for Peplography and DLIM is single refocused scattering light field image. Fig. 5 shows the reference images, the input color light field images, Peplography reconstructed images and DLIM reconstructed images in (a)-(d). It reveals that the self-luminous targets cannot be recognized from Peplography-reconstructed images but are well observed from DLIM-reconstructed images as seen in (d) of Fig. 5. To quantitively evaluate the imaging capacity of proposed method, the two most commonly used image evaluation methods, SSIM and PSNR are applied. The reference images are captured with the same light field imaging system without fog. The SSIM of Peplography reconstructed images for four groups of self-luminous targets of 'T', 'H', 'U' and '2' are 0.002, 0.015, 0.029 and 0.048, whereas the DLIM reconstructed images have SSIM values of 0.049, 0.105, 0.266 and 0.363, respectively. And the PSNRs are 9.837, 14.766, 16.950 and 18.702 for Peplography and 18.433, 19.382, 20.181 and 21.276 for DLIM, respectively.

### 1) The effect of anisotropic factor to diffusion light field approximation

As mentioned before, the scattering media should be slightly anisotropic to allow the neglect of high-order spherical harmonic expansion of radiances [36, 37]. To test the effect of anisotropic factor to diffusion light field approximation, we carry out the rendering of scattering light field images by setting different anisotropic factors, in which the light field images are then captured with varying isotropic factor from 0 to 0.8. As seen in Fig. 7, both scattering light field images and reconstructed light field images with DLIM are shown. In the above row, it is obviously that more scattering photons are concentrated in a small radius region when the isotropic factor is 0.8, which means that the photon distribution is highly anisotropic. The distribution radius of the scattering photon region is expanded as the anisotropic factor becomes smaller. Although the same scattering strength set along different anisotropic factors, the target is successfully reconstructed only

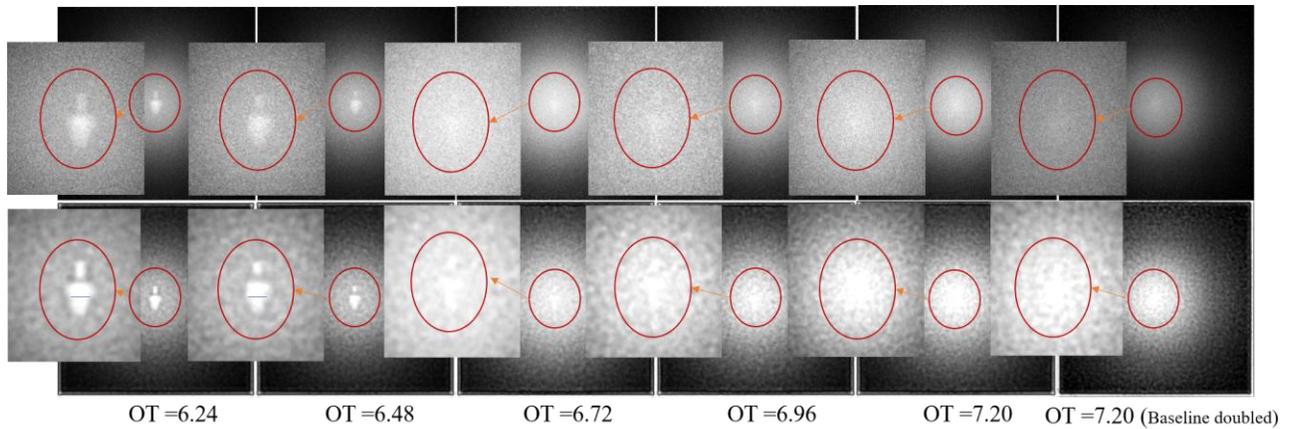

Fig. 8. Experimental results for verification of reconstruction capacity of DLIM relating to optical thickness.

### C. Evaluating the proposed DLIM through simulation experimentations and descriptive factor analysis

Here, several critical factors are analyzed with simulation experiments to investigate their effect to diffuse light field imaging framework. We carry out simulation experiments with a physical-based rendering platform Mitsuba that can simulate scattering environments with the Monte Carlo method by setting different scattering factors of media and the objects are set as self-luminous [43, 44]. As shown in Fig. 6(a), the simulation diagram is set as same with practical experiments of the fog chamber. The target emerges into the scattering media and the cameras are outside of the media. Since the scattering media is static, we capture perspective images of the light field by moving cameras in both horizontal and vertical directions with identical displacements. Nine positions are sampled for each of the horizontal and vertical directions, so that the scattering light field images contain a total of 81 perspectives. As shown in Fig. 6(b), the rendering environment consists of three components: the target 'Vase', scattering media with adjustable scattering coefficients, and the camera. Several critical factors consist of anisotropic factor, optical thickness and three-dimensional imaging capacity are investigated, respectively.

when the anisotropic factor is equal to 0, indicating that the reconstruction capability of scattering light field will be maximum for isotropic scattering media. Intuitively, isotropic media will scatter photons more broadly and uniformly, the scattering process becomes random-walk in the scattering volume and the diffusion approximation is more accurate, which is the fundamental of diffuse light field imaging framework as well.

### 2) Reconstruction capacity of DLIM in relation to optical thickness

The reconstruction capacity relating to optical thickness for proposed DLIM is investigated as well. To accurately control the optical thickness, we use Mitsuba to render the scattering images with continuously increased scattering coefficient and keep the other parameters fixed. The optical thickness (OT) is increased from 6.24 to 7.20, the scattering coefficient is from 0.026 to 0.3 and thickness of scattering media is 240. As seen in Fig. 8, the refocused light field images and the reconstructed images are in upper and lower rows. The object is directly visible in refocused scattering image when OT equals 6.24, and the reconstructed image has a more concentrated energy distribution in the object area and sharper edge. When OT = 6.48, the object is still visible in the refocused scattering image, but only the edge is visible in the reconstructed image. At OT = 6.72, OT = 6.96 and OT = 7.20, the object is not directly visible in the refused scattering images, but most of the object information can be recognized in the reconstructed images. Although the shape of object cannot be recognized in the reconstructed image of OT = 7.2, the energy concentration in object area is obviously. Note that the reconstructed object has an uneven energy distribution. For example, the most convex region of object highlighted by the blue line in Fig. 8 when OT = 6.24 and 6.48 has a higher intensity and the neck regions of object are absent in all images. There are several reasons for this phenomenon. First, the bi-directional scattering distribution function (BSDF) and depth difference of object surface are not considered in reconstruction. Secondly, the capture window composed by camera array is limited. That is, the BSDF of object causes uneven light detection for each point of the object source through the limited capture window. Since most of light for some regions, e. g. convex region, are captured by capture window, and some region, such as neck region, scatter light towards directions out of captured window. The depth difference will cause different attenuation ratio for each pixel of object image and the convex area of object 'Vase' has shortest depth implying smallest attenuation ratio. Briefly, above factors caused the energy unbalance in reconstructed images. From this group of experiments, we also deduce that the capture window of light field is also impactful to reconstruction results, since different light flux for different area of object can be captured finally depending on capture window. Particularly when OT is large, this impact is more prominent, since most regions of object even has no ballistic photons are collected by cameras, which results in no diffuse sources are constructed in refocused image (non-source diffusion: similar photon distribution probability for each pixel). This can be alleviated by enlarging capture window and capture ballistic photons from more directions. It has been confirmed by experiments. The last column of Fig. 8 shows the results by double the baseline of camera array and other parameters are fixed, object can be reconstructed again accompanied with more complete information even the neck area comparing to previous reconstruction results.

### 3) Verification for three-dimensional imaging capacity of DLIM

The experimental results investigating the three-dimensional imaging performance of the proposed DLIM are shown in Fig. 9. The 1st row shows the clear refocused light field images without scattering, containing three objects located at (10, 10, 150), (10, 0 200) and (0, 0 240), respectively. The 2nd and 3rd rows show the refocused scattering images and the reconstructed images corresponding to the images in 1st row. The 1st, 2nd and 3rd columns show all images refocused at distance of 150, 200 and 240, respectively.  For refocused scattering images, the object image gradually disappears as the distance of the object increases until it disappears completely at a refocusing depth of 240 due to the optical thickness increases. The reconstructed images show obvious rebuilt object when reconstruction depth is 150 and 200. Although the objects at the depths of 200 and 240 are reconstructed simultaneously and cause interference at a reconstruction distance of 240, the ability to recover signal from a highly scattering image with no visible source to an energy-converged image with object source distribution is confirmed. For the refocus of scattering light field, we find that the refocus of front object has weak interference, while the refocus of rear object causes stronger interference. This is plausible because the rear object has a lower signal than the front object due to the stronger scattering.

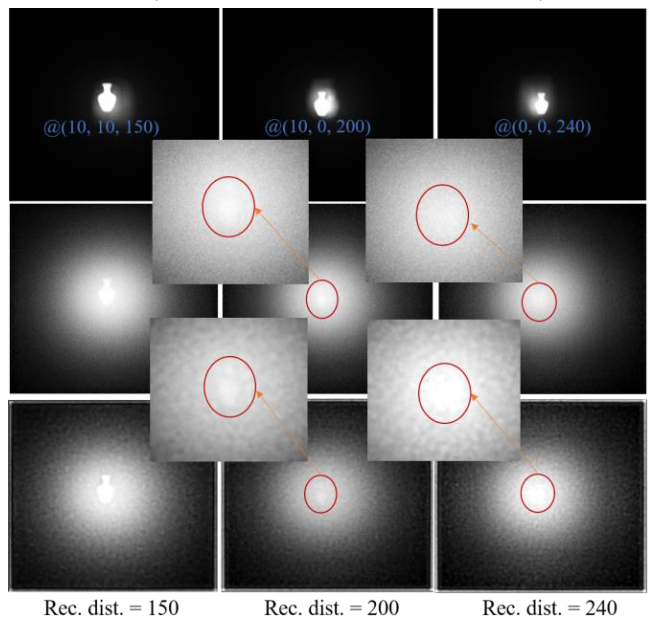

Fig. 9. Experimental results for verification of three-dimensional imaging capacity of DLIM.

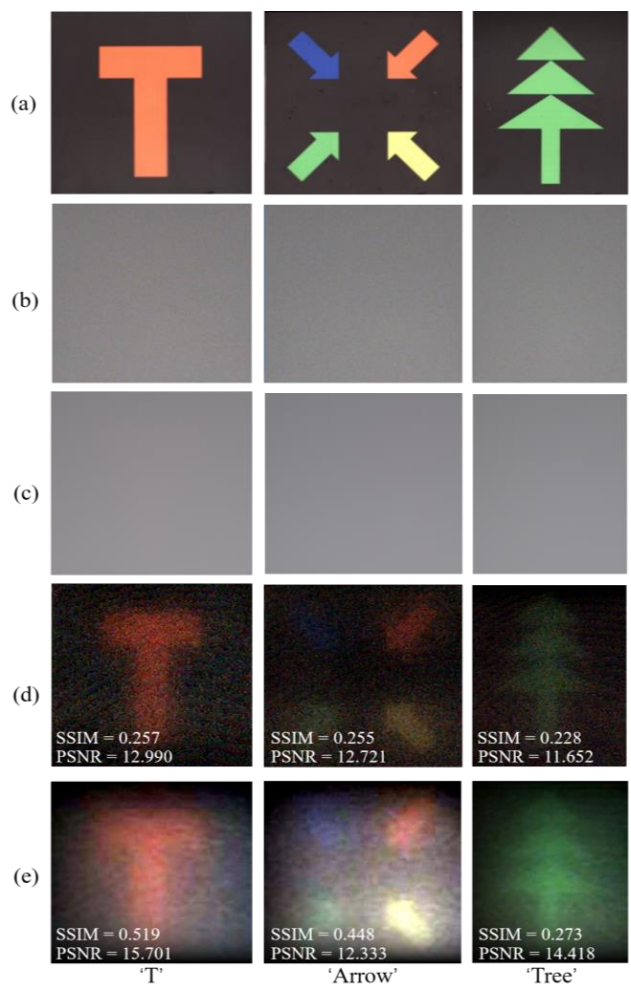

Fig. 10. Experimental comparison between Peplography and DLIM regarding passive-luminous scattering light field images.

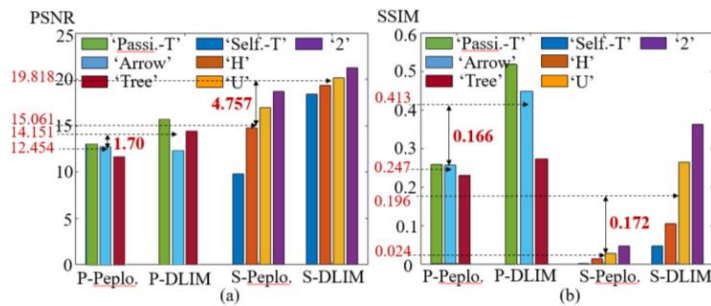

Fig. 11. Statistic analysis of reconstruction quality with respect to PSNR and SSIM for Peplography and DLIM in dealing with both self- and passive-luminous objects. (a) PSNR statistic result. (b) SSIM statistic result.

*D. Verification of DLIM for passive-luminous objects through super dense fog considering backward scattering modeling*

Moreover, three groups of experiments are conducted for passive-luminous targets 'T', 'Tree', and 'Arrow' to verify the effectiveness of the proposed method for scattering imaging in dealing with backscattering photons as shown in Fig. 10. These experiments are also carried out under the fog chamber. With our measurement equipment, the optical thickness in this case is about 9, which is about 3 times visibility by using optical thickness dividing $|log 0.05|$ [41]. As seen in Fig. 10, the images in the first row are ground truth, the second and third rows show the perspective scattering images and refocused scattering light field images. Due to the backward scattering photon the SNR of scattering images are super low. By the way, different from self-luminous scattering images with color distortion as seen in Fig.4, as the passive-luminous light source is white color, the scattering images are all looks gray color with random noise. The images in the second and last rows are the reconstructed images from scattering light field images using Peplography and DLIM, respectively. The experimental results reveal that the SSIM of Peplography-reconstructed images for three passive-luminous targets of 'T', 'Arrow', and 'Tree' are 0.257, 0.255, and 0.228, whereas the DLIM reconstructed images are 0.519, 0.448, and 0.273, respectively. And the PSNRs are 12.990, 12.721, and 11.652 for Peplography and 15.701, 12.333, and 14.418 for DLIM, which indicates the higher performance of DLIM than Peplography in terms of SSIM and PSNR.

In summary, the results of the statistical analysis for Peplography, DLIM, and SLIM are shown in Fig. 11. For passive-luminous targets, the average PSNR of Peplography and SLIM are 12.454 and 14.151, which indicates higher 1.70 PSNR gained by SLIM than Peplography; the average SSIM are 0.247 and 0.413, which means higher 0.166 SSIM gained by SLIM than Peplography. For self-luminous targets, the average PSNR of Peplography and DLIM are 15.061 and 19.818, which indicates a higher 4.757 PSNR gained by DLIM than Peplography; the average SSIM are 0.024 and 0.196, which means higher 0.172 SSIM gained by DLIM than Peplography. Furthermore, the state-of-the-art single image dehazing methods have been also compared with proposed DLIM method. To guarantee the fair of comparison, all methods use single refocused light field image as input. And the experimental results reveal that the proposed method still gains highest PSNR no matter for the self-luminous or passive-luminous cases. Detailed experiment implementation and results can be found in Supplementary. Noting that, those compared dehazing methods aim to make the weak fog image to be clear, whereas the DLIM focus on imaging through highly scattering media.

## IV. DISCUSSIONS AND CONCLUSIONS

The currently proposed model assumes that the scattering medium is homogeneous. In practice, materials can have inhomogeneous scattering and absorption coefficients as non-uniform density distribution of volume particles. The proposed technique could potentially be extended to account for the non-uniform density of scattering media by modeling the variance under average density when performing another convolution to the diffuse kernel [45]. In the case of serious anisotropy of media, then the diffusion approximation will be invalid. In this case, only the part of radiate modeling for scattering light field model can be applicable and the solution of RTE cannot be simplified as $P_1$ approximation. High-order $P_N$ approximation can be used to obtain a solution for 4-D radiate kernel.

Optical imaging overcoming the scattering effect is challenging but significant for many fields. In this paper, a diffuse light field model is proposed to construct a novel light field imaging framework for optical imaging through highly scattering media. Moreover, by combining the classical J-M model, a complete scattering light field model can be constructed, which is a fundamental theoretical work that will feed more scattering imaging algorithms. Extensive experiments have been conducted to confirm the superior imaging capacity of the proposed methods compared to state-of-the-art scattering imaging methods for both of passive-luminous and self-luminous targets. To the best of our knowledge, this is the first physically-aware scattering light field imaging model, which might extend the light field imaging framework into scattering imaging area.

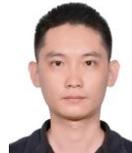

**Hongkun Cao** received the Ph. D. degree in electronics engineering from Kwangwoon University, Seoul, South Korea, in 2020.
From 2020-2022, he was a Postdoctoral Fellow with the Shenzhen International Graduate School, Tsinghua University. Since 2022, he has been with Peng Cheng Laboratory, where he is a research associate. His current research interests include computational imaging, light filed holography and computer rendering. He was awarded as Shenzhen Overseas High-level Talents (Category C)  in 2020.

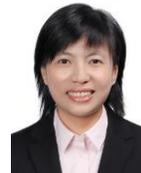

**Xin Jin** (S'03–M'09-SM'11) received the M.S. degree in communication and information system and the Ph.D. degree in information and communication engineering, both from Huazhong University of Science and Technology, Wuhan, China, in 2002 and 2005, respectively.
From 2006 to 2008, she was a Postdoctoral Fellow with The Chinese University of Hong Kong. From 2008 to 2012, she was a Visiting Lecturer with Waseda University, Fukuoka, Japan. Since Mar. 2012, she has been with Shenzhen International Graduate School, Tsinghua University, China, where she is currently a professor. Her current research interests include computational imaging and power-constrained video processing and compression. She has published over 150 conference and journal papers.
  Dr. Jin is an IEEE Senior Member, and a member of SPIE and ACM. She is the Distinguished Professor of Pengcheng Scholar. She received Second prize of National Science and Technology Progress Award in 2016, the first prize of Guangdong science and technology award in 2015 and ISOCC Best Paper Award in 2010.

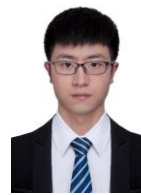

**Junjie Wei** is currently working toward the Ph.D. degree in the Control Science and Engineering with Shenzhen International Graduate School, Tsinghua University, China. His research interests include holographic display, computational imaging, light filed holography.

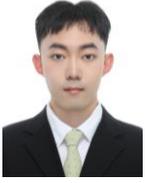
**Yihui Fan** is currently working toward the Ph.D. degree in the Control Science and Engineering with Shenzhen International Graduate School, Tsinghua University, China. His research interests include the development of new systems and algorithms for solving problems in scattered light field sampling and scattering imaging.

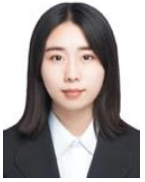
**Dongyu Du** is currently working toward the Ph.D. degree in the Control Science and Engineering at Tsinghua University Automation Department, China. Her research interests include the development of new systems and algorithms for solving problems in scattering imaging, non-line-of-sight imaging and single-photon imaging.